\documentclass[prd,aps,showpacs,nofootinbib]{revtex4}%
\usepackage{amsmath}
\usepackage{amsfonts}
\usepackage{amssymb}
\usepackage[dvips]{epsfig}
\usepackage[dvips]{graphicx}%
\providecommand{\U}[1]{\protect\rule{.1in}{.1in}}
\newcommand{\be}{\begin{equation}}
\newcommand{\ee}{\end{equation}}
\newcommand{\bq}{\begin{eqnarray}}
\newcommand{\eq}{\end{eqnarray}}
\begin{document}
\title{Classical solutions for the Carroll-Field-Jackiw-Proca electrodynamics }
\author{Rodolfo Casana, Manoel M. Ferreira Jr, Carlos E. H. Santos}
\affiliation{Departamento de F\'{\i}sica, Universidade Federal do Maranh\~{a}o (UFMA),
Campus Universit\'{a}rio do Bacanga, S\~{a}o Lu\'{\i}s-MA, 65085-580 - Brasil}

\begin{abstract}
In the present work, we investigate classical solutions of the
Maxwell-Carroll-Field-Jackiw-Proca (MCFJP) electrodynamics for the cases a
purely timelike and spacelike Lorentz-violating (LV) background. Starting from
the MCFJP Lagrangian and the associated wave equations written for the
potential four-vector, the tensor form of the Green function is achieved. In
the timelike case, the components of the stationary Green function are
explicitly written. The classical solutions for the electric and magnetic
field strengths are then evaluated, being observed that the electric sector is
not modified by the LV background, keeping the Maxwell-Proca behavior. The
magnetic field associated with a charge in uniform motion presents an
oscillating behavior that also provides an oscillating MCFJ solution (in the
limit of a vanishing Proca mass), but does not recover the Maxwell-Proca
solution in the limit of vanishing background. In the spacelike case, the
stationary Green function is written and also explicitly carried out in the
regime of a small background. The electric and magnetic fields reveal to
possess an exponentially decaying behavior, that recover the Maxwell-Proca solutions.

\end{abstract}

\maketitle

\section{Introduction}

Lorentz covariance is regarded as one of the fundamental symmetries of nature.
The rise and establishment of such a symmetry begun with the advent of Special
Theory of Relativity, being after incorporated as a key feature of the modern
Field Theories. Nowadays, Lorentz covariance pervades all known physical
interactions, having the status of a cornerstone of modern physical theories.
Investigations concerning LV are mostly conducted under the framework of the
SME - Standard Model Extension, developed by Colladay \& Kostelecky
\cite{Colladay}. The Standard Model Extension (SME) is a broader version of
the usual Standard Model that embraces all Lorentz-violating (LV) coefficients
(generated as vacuum expectation values of tensor quantities belonging to an
underlying theory at the Planck scale) that yield Lorentz scalars (as tensor
contractions) in the observer frame. Such coefficients govern Lorentz
violation in the particle frame, where they are seen as sets of independent
numbers, whereas they work out as genuine tensor in the observer frame. One
strong motivation to study the SME refers to the desire to get information on
the Planck scale physics, where Lorentz violation may be allowed. This
possible breaking is an important information for the development of quantum
gravity theory. This kind of idea was supported by the demonstration
concerning the possibility of Lorentz and CPT spontaneous breaking in the
context of string theory \cite{Samuel}. In this scenario, a minuscule Lorentz
violation at a lower energy scale (scrutinized into the framework of the SME)
is to be read as a remanent effect of spontaneous Lorentz violation at Planck
scale. Nowadays, Lorentz violation has been investigated in many different
systems and purposes \cite{General}, involving also fermions \cite{Fermions},
CPT and Lorentz-violating probing tests \cite{Tests}, topological phases
\cite{Phases}, radiative corrections \cite{Radiative} and the gauge sector
\cite{Photons,Cerenkov}.

Lorentz violation in the photon sector of the SME has been much investigated
in literature with basically a twofold purpose: the determination of new
electromagnetic effects induced by the Lorentz-violating coefficients and the
imposition of stringent upper bounds on the LV coefficients that constrain the
magnitude of Lorentz breaking. The pioneer investigation of LV effects on
classical electromagnetism was performed by Carroll-Field-Jackiw \cite{CFJ},
who studied the Maxwell electrodynamics in the presence of the assigned
Carroll-Field-Jackiw term $\left(  \epsilon^{\beta\alpha\rho\lambda}V_{\beta
}A_{\alpha}F_{\rho\lambda}\right)  $, with $V_{\beta}$ standing for the LV
fixed background. This Lorentz and CPT-odd term leads to a gauge invariant
theory - Maxwell-Carroll-Field-Jackiw (MCFJ) electrodynamics, which is causal,
stable and unitarity only for a purely spacelike background \cite{Adam}. The
photon sector of SME is composed by the Carroll-Field-Jackiw term and another
Lorentz-violating term $\left(  K^{\beta\alpha\rho\lambda}F_{\beta\alpha
}F_{\rho\lambda}\right)  $, which is CPT-even, with $K^{\beta\alpha\rho
\lambda}$ being the LV tensor coefficient. An interesting study on the
electrostatic and magnetostatic features associated with this term was
performed by Bailey \& Kostelecky \cite{Bailey}, which have used the Green
function techniques to obtain the classical solutions for the 4-potential
vector in vacuum and in a material medium. Such solutions revealed the
interconnection existing between the electric and magnetic sectors in this
theory. Other studies involving LV in electrodynamical models are also known
in literature \cite{Electrodynamic}.

Lorentz violation in the presence of the Higgs sector has also been examined,
with the purpose of establishing upper bounds on the associate breaking
parameters and studying the Nambu-Goldstone modes \cite{Higgs}. An
investigation of the Higgs sector in the context of the MCFJ model was
accomplished as well \cite{Baeta}. The resulting Mawell-Carroll-Field-Jackiw
electrodynamics with the Proca mass term - Maxwell-Carroll-Field-Jackiw-Proca
(MCFJ-Proca) electrodynamics, has had its consistency examined, exhibiting an
outcome similar to that of the MCFJ model, that is, the causality and
unitarity are assured for a purely spacelike background whereas are spoiled
for a purely timelike background. It is still worthy to mention that several
properties of the MCFJ electrodynamics were already addressed in a
low-dimension space-time. Indeed, such model was properly undergone to a
dimensional reduction to (1+2) dimensions, yielding a Maxwell-Chern-Simons
(MCS)\ electrodynamics coupled to a massless Klein-Gordon field (stemming from
the previous $A^{\left(  3\right)  }$ component) and a two-dimensional LV
background. The consistency of this planar model was examined, revealing a
model totally causal, stable and unitary \cite{Belich}. The static classical
solutions of this planar model were determined for a point-like charge
\cite{Manojr1}, revealing the background effects on the electric and magnetic
sectors of the MCS electrodynamics. A similar study was also performed for the
case of the Maxwell-Carroll-Field-Jackiw electrodynamics with Higgs sector.
Such a model was dimensionally reduced to (1+2) dimension as well, having its
consistency and classical solutions properly examined \cite{Manojr2}.

Such detailed investigations on the classical solutions in (1+2) dimensions
have not counterpart in the original models \cite{CFJ},\cite{Baeta}, defined
in (1+3) dimensions. Hence, the purpose of the present work is to study the
classical solutions of the MCFJ and MCFJ-Proca models for both purely timelike
and purely spacelike backgrounds, for static and stationary sources. The
starting point in both cases is the evaluation of the Green function for the
tensor equation for the four-potential $A^{\mu}$, which provides Fourier
expressions for the scalar and vector potential in the momentum space. The
Fourier transforms of such relations lead to the classical solutions for these
potentials, which yield the solutions for the field strengths. In the purely
timelike case, the stationary Green function is evaluated. It is observed that
the electric field presents an exponentially decaying behavior, independent of
the background, equal to the usual Maxwell-Proca result. The magnetic field is
null for a static charge and exhibits an intricate oscillating behavior for a
stationary moving charge. The limit of a vanishing Proca mass yields the
stationary MCFJ solutions. For the case of a purely spacelike background, no
exact Fourier transforms for the potentials are obtained. The integrals are
then performed under the approximation of a small background ($\mathbf{v}%
^{2}\ll M_{A}^{2}),$ and the stationary Green function carried out. The
electric and magnetic field strengths, evaluated at the $\mathbf{v}^{2}%
$-order, exhibit an exponentially decaying behavior. In the limit of a
vanishing background, the results recover the usual Maxwell-Proca solution.

This paper is outlined as follows. In Sec. II, it is shown a brief
presentation of the basic aspects of the classical MCFJ-Proca model, including
equations of motion, energy-momentum tensor, and classical wave equations. In
Sec. III, we proceed with the evaluation of the Green function associated with
the tensor equation for the four-potential. The stationary Green function is
computed explicitly. Expressions for the scalar and vector potentials are
derived, which provides explicit solutions for the electric and magnetic field
strengths. This is done both for a purely timelike and spacelike background configuration.

\section{The Carroll-Field-Jackiw Electrodynamics with Proca mass}

The starting point is the Carroll-Field-Jackiw-Proca Lagrangian, written in
(1+3) dimensions:
\begin{equation}
\mathcal{L}=-\frac{1}{4}F_{\alpha\nu}F^{\alpha\nu}+\frac{1}{2}M_{A}%
^{2}A_{\alpha}A^{\alpha}-\frac{1}{4}\varepsilon^{\beta\alpha\rho\lambda
}V_{\beta}A_{\alpha}F_{\rho\lambda}+J^{\alpha}A_{\alpha}, \label{L1}%
\end{equation}
with $V^{\alpha}=\left(  \text{v}_{0},\text{\textbf{v}}\right)  $ being the
fixed background responsible for Lorentz-violation in the gauge sector. Such a
model was by first considered in ref. \cite{Baeta}, where the Proca mass stems
from a Higgs scalar sector. The gauge propagator was evaluated and its
consistency was analyzed. It was then shown that this model is unitary just
for a spacelike background while it presents ghost states for a timelike or
lightlike background. The Euler-Lagrange equation leads to the modified
Maxwell equation,
\begin{align}
\partial_{\nu}F^{\nu\alpha}+M_{A}^{2}A^{\alpha}+V_{\beta}\tilde{F}%
^{^{\alpha\beta}}  &  =-J^{\alpha},\label{motion1}\\
\partial_{\alpha}\tilde{F}^{^{\alpha\beta}}  &  =0, \label{motion2}%
\end{align}
where $\tilde{F}^{\alpha\beta}=\displaystyle\frac{1}{2}\epsilon^{\alpha
\beta\mu\nu}F_{\mu\nu}$ is the dual tensor, with the convention $\epsilon
^{0123}=+1$. From eq. (\ref{motion1}), we obtain $M_{A}^{2}\partial_{\alpha
}A^{\alpha}=-\partial_{\alpha}J^{\alpha}.$ Considering current conservation
($\partial_{\alpha}J^{\alpha}=0),$ the Lorentz gauge $(\partial_{\alpha
}A^{\alpha}=0)$ appears as an implied condition. The energy and momentum
storaged by the electromagnetic field may be taken by the energy-momentum
tensor:
\[
\Theta^{\alpha\beta}=-F^{\alpha\nu}F_{\text{ }\nu}^{\beta}+\frac{1}%
{4}g^{\alpha\beta}F_{\mu\nu}F^{\mu\nu}+\frac{1}{4}\epsilon^{\alpha\nu
\lambda\rho}V^{\beta}A_{\nu}F_{\lambda\rho}-\frac{1}{2}g^{\alpha\beta}%
M_{A}^{2}A_{\lambda}A^{\lambda}%
\]
Once this theory is invariant under space-time translations, the
energy-momentum tensor is conserved ($\partial_{\alpha}\Theta^{\alpha\beta
}=0)$ in the absence of sources. This tensor can not be turned symmetric as a
consequence of Lorentz-violation. The energy density is the written as
\[
\Theta^{00}=\frac{1}{2}[\mathbf{E}^{2}+\mathbf{B}^{2}+\mathbf{v}_{0}%
\mathbf{B}\cdot\mathbf{A-}M_{A}^{2}(A_{0}^{2}-\mathbf{A\cdot A)].}%
\]
This expression reveals that the energy is not positive definite due to the
term $\mathbf{v}_{0}\mathbf{B}\cdot\mathbf{A}$, which may be negative.

The motion equations (\ref{motion1}, \ref{motion2}) are explicitly written as
the modified Maxwell equations:\textbf{ }%
\begin{align}
\nabla\cdot\mathbf{E}+M_{A}^{2}A^{0}+\mathbf{v}\cdot\mathbf{B} &
=-\rho,\label{M1}\\[0.2cm]
\nabla\times\mathbf{B}-\partial_{t}\mathbf{E-v}\times\mathbf{E} &
=-\mathbf{v}_{0}\mathbf{B}-M_{A}^{2}\mathbf{A}-\mathbf{j},\label{M2}\\[0.2cm]
\nabla\times\mathbf{E} &  =-\partial_{t}\mathbf{B},\label{M3}\\[0.2cm]
\nabla\cdot\mathbf{B} &  =0.\label{M4}%
\end{align}
Manipulating these relations, wave equations for field strengths are readily
attained:%
\begin{align}
\left(  \square+M_{A}^{2}\right)  \mathbf{B}+\mathbf{v}_{0}\left(
\nabla\times\mathbf{B}\right)   &  =\nabla\times\left(  \mathbf{v}%
\times\mathbf{E}\right)  -\nabla\times\mathbf{j},\label{B1}\\[0.2cm]
\left(  \square+M_{A}^{2}\right)  \mathbf{E}+\partial_{t}\left(
\mathbf{v}\times\mathbf{E}\right)   &  =\nabla\left(  \mathbf{v}%
\cdot\mathbf{B}\right)  +\mathbf{v}_{0}\partial_{t}\mathbf{B}+\nabla
\rho+\partial_{t}\mathbf{j}.\label{E1}%
\end{align}

Wave equations can be also written for the four-potential:%
\begin{equation}
\left[  \left(  \square+M_{A}^{2}\right)  g^{\mu\nu}+\epsilon^{\mu\nu
\kappa\lambda}V_{\kappa}\partial_{\lambda}\right]  A_{\nu}=-J^{\mu
},\label{A_motion}%
\end{equation}
whose scalar and vector components are:%
\begin{align}
\left(  \square+M_{A}^{2}\right)  A^{0}+\mathbf{v}\cdot\mathbf{B} &
=-\rho,\\[0.2cm]
\left(  \square+M_{A}^{2}\right)  \mathbf{A}+\mathbf{v}_{0}\mathbf{B-v}%
\times\mathbf{E} &  =-\mathbf{j}.
\end{align}
These equations reveal a remarkable feature of the CFJ electrodynamics: the
electric and magnetic sectors become entwined for the case the background
presents a non-null space component $\left(  \mathbf{v\neq0}\right)  $. In
this situation, the magnetic field strength contributes for the determination
of the scalar potential while the electric field strength affects the vector
potential solution. This means that charge and current densities both
contribute to electric and magnetic field solutions, so that a static charge
originates both electric and magnetic field strengths and a stationary current
yields both magnetic and electric fields. A similar mixing of the electric and
magnetic sectors is also reported in the context of the electrodynamics
related to the term $K^{\beta\alpha\rho\lambda}F_{\beta\alpha}F_{\rho\lambda}$
(see ref. \cite{Bailey}). For a purely timelike background, $V^{\mu
}=(\mathbf{v}_{0},\mathbf{0}),$ the potential equations decouple and the
sector entanglement ceases, recovering the usual uncoupled electromagnetic behavior.

\section{Solution by the Green Method}

A complete solution for the potentials can be obtained by the Green method.
The implementation of the Green method begins by writing the 4-potential and
the 4-current as Fourier transforms in momentum space:
\begin{align}
A_{\mu}\left(  r\right)   &  =\int\frac{d^{4}p}{\left(  2\pi\right)  ^{4}%
}\tilde{A}_{\mu}\left(  p\right)  \exp\left(  -ip\cdot r\right)  ,\\
J^{\mu}\left(  r\right)   &  =\int\frac{d^{4}p}{\left(  2\pi\right)  ^{4}%
}\tilde{J}^{\mu}\left(  p\right)  \exp\left(  -ip\cdot r\right)  ,
\end{align}
Such expressions must be replaced in eq. (\ref{A_motion}), providing:
\begin{equation}
D^{\mu\nu}\tilde{A}_{\nu}\left(  p\right)  =\tilde{J}^{\mu}\left(  p\right)  ,
\end{equation}
The tensor operator $D^{\nu\beta}$ is fully written as
\begin{equation}
D^{\mu\nu}=\left(  p^{2}-M_{A}^{2}\right)  g^{\mu\nu}+i\varepsilon^{\mu
\nu\alpha\kappa}V_{\alpha}p_{\kappa},
\end{equation}
Its determinant is
\begin{equation}
D=\left(  p^{2}-M_{A}^{2}\right)  \left[  \left(  p\cdot V\right)  ^{2}%
-p^{2}V^{2}-\left(  p^{2}-M_{A}^{2}\right)  ^{2}\right]  ,
\end{equation}
whose zeros play a relevant role in the investigation of the spectrum of the
theory, as suitably shown in ref. \cite{Baeta}.

The solution for the 4-potential can be constructed in terms of the inverse
tensor operator, $\left[  D_{\mu\nu}\right]  ^{-1},$ that is:
\begin{equation}
\tilde{A}_{\mu}\left(  p\right)  =\left[  D^{-1}\right]  _{\mu\nu}\tilde
{J}^{\nu}\left(  p\right)  . \label{A1}%
\end{equation}

\begin{equation}
\left[  D^{-1}\right]  _{\mu\nu}=\frac{1}{Q}\left\{  \left(  p^{2}-M_{A}%
^{2}\right)  g_{\mu\nu}+\frac{V^{2}}{p^{2}-M_{A}^{2}}p_{\mu}p_{\nu}%
-i\epsilon_{\mu\nu\rho\sigma}V^{\rho}p^{\sigma}+\frac{{p}^{2}}{p^{2}-M_{A}%
^{2}}V_{\mu}V_{\nu}-\frac{\left(  V\cdot p\right)  }{p^{2}-M_{A}^{2}}\left(
V_{\mu}p_{\nu}+V_{\nu}p_{\mu}\right)  \right\}  , \label{Mat1}%
\end{equation}
with%
\begin{equation}
Q=\left(  {p}^{2}-M_{A}^{2}\right)  ^{2}+V^{2}p^{2}-\left(  V\cdot p\right)
^{2}.
\end{equation}
It is important to point out that the inverse tensor $\left[  D_{\mu\nu
}\right]  ^{-1}$ is not equal to the photon sector propagator obtained in ref.
\cite{Baeta}. Now, we shall particularize the tensor $D^{\nu\beta}$ and the
associated solutions for the cases the LV background is purely timelike or
purely spacelike.

\subsection{Solution for a purely timelike background}

For a purely timelike background, $V^{\alpha}=(\mathbf{v}_{0},\mathbf{0}),$
the components of the inverse tensor operator, $\left[  D_{\nu\beta}\right]
^{-1}$, takes the form%
\begin{align}
\left[  D^{-1}\right]  _{00}\left(  p\right)   &  =\frac{1}{{p}^{2}-M_{A}^{2}%
},~\ \ \ \ \ \ \left[  D^{-1}\right]  _{0k}=\left[  D^{-1}\right]
_{k0}=0,\label{D_inverse1}\\
\left[  D^{-1}\right]  _{jk}\left(  p\right)   &  =\frac{1}{\Delta}\left\{
-\left(  p^{2}-M_{A}^{2}\right)  \delta_{jk}+\frac{\mathbf{v}_{0}^{2}}%
{p^{2}-M_{A}^{2}}p_{j}p_{k}-i\epsilon_{jkl}\mathbf{v}_{0}p_{l}\right\}  ,
\label{D_inverse1B}%
\end{align}
where $\epsilon_{123}=1$ and $\Delta$ is
\begin{equation}
\Delta=\left(  {p}^{2}-M_{A}^{2}\right)  ^{2}-\mathbf{v}_{0}^{2}\mathbf{p}%
^{2}.
\end{equation}

We are interested in stationary solution of the four-potential, thus, setting
$p_{0}=0$ the components of the tensor $\left[  D_{\nu\beta}\right]  ^{-1}$
read as%
\begin{align}
\left[  \bar{D}^{-1}\right]  _{00}\left(  \mathbf{p}\right)   &  =-\frac
{1}{\mathbf{p}^{2}+M_{A}^{2}}~\ \ \ ,~\ \ \ \ \left[  \bar{D}^{-1}\right]
_{0k}=\left[  \bar{D}^{-1}\right]  _{k0}=0,\\
\left[  \bar{D}^{-1}\right]  _{jk}\left(  \mathbf{p}\right)   &  =\frac
{1}{\bar{\Delta}}\left\{  \left(  \mathbf{p}^{2}+M_{A}^{2}\right)  \delta
_{jk}-\frac{\mathbf{v}_{0}^{2}}{\mathbf{p}^{2}+M_{A}^{2}}p_{j}p_{k}%
-i\epsilon_{jkl}\mathbf{v}_{0}p_{l}\right\}  ,
\end{align}
with%
\begin{equation}
\bar{\Delta}=\left(  \mathbf{p}^{2}+M_{A}^{2}\right)  ^{2}-\mathbf{v}_{0}%
^{2}\mathbf{p}^{2}.
\end{equation}

We should now write a general expression for the four-potential by the Green
method taking a non-null current density. In this sense, the four-potential is
read as
\begin{equation}
A_{\mu}\left(  \mathbf{r}\right)  =%
{\displaystyle\int}
d^{3}\mathbf{r}^{\prime}G_{\mu\nu}\left(  \mathbf{r}-\mathbf{r}^{\prime
}\right)  J^{\nu}\left(  \mathbf{r}^{\prime}\right)  , \label{A_pot}%
\end{equation}
where $G_{\mu\nu}\left(  \mathbf{r}-\mathbf{r}^{\prime}\right)  $ is the
Green's functions here written in terms of the inverse tensor $\left[  \bar
{D}^{-1}\right]  _{\mu\nu}$:
\begin{equation}
G_{\mu\nu}\left(  \mathbf{r}-\mathbf{r}^{\prime}\right)  =%
{\displaystyle\int}
\frac{d^{3}\mathbf{p}}{\left(  2\pi\right)  ^{3}}\left[  \bar{D}^{-1}\right]
_{\mu\nu}\left(  \mathbf{p}\right)  ~e^{i\mathbf{p}\cdot\left(  \mathbf{r}%
-\mathbf{r}^{\prime}\right)  }.
\end{equation}
From the matrix $\left[  \bar{D}^{-1}\right]  _{\mu\nu}\left(  \mathbf{p}%
\right)  ,$ we can write straightforwardly the components of the Green
function in terms of Fourier integrals:
\begin{align}
G_{00}\left(  \mathbf{R}\right)   &  =-I_{0}\left(  \mathbf{R}\right)
,\label{G1}\\
G_{0k}\left(  \mathbf{R}\right)   &  =G_{k0}\left(  \mathbf{R}\right)
=0,\label{G2}\\
G_{jk}\left(  \mathbf{R}\right)   &  =\delta_{jk}I_{1}\left(  \mathbf{R}%
\right)  +\mathbf{v}_{0}^{2}\partial_{j}\partial_{k}I_{2}\left(
\mathbf{R}\right)  -\mathbf{v}_{0}\epsilon_{jkl}\partial_{l}I_{3}\left(
\mathbf{R}\right)  , \label{G3}%
\end{align}
where the integrals $I_{i}\left(  \mathbf{R}\right)  $ are defined bellow:
\begin{align}
I_{0}\left(  \mathbf{R}\right)   &  =%
{\displaystyle\int}
\frac{d^{3}\mathbf{p}}{\left(  2\pi\right)  ^{3}}\frac{~e^{i\mathbf{p}%
\cdot\mathbf{R}}}{\mathbf{p}^{2}+M_{A}^{2}}=\frac{1}{4\pi}\frac{e^{-M_{A}R}%
}{R},\\
I_{1}\left(  \mathbf{R}\right)   &  =%
{\displaystyle\int}
\frac{d^{3}\mathbf{p}}{\left(  2\pi\right)  ^{3}}\frac{\left(  \mathbf{p}%
^{2}+M_{A}^{2}\right)  ~e^{i\mathbf{p}\cdot\mathbf{R}}}{\left(  \mathbf{p}%
^{2}-m_{+}^{2}\right)  \left(  \mathbf{p}^{2}-m_{-}^{2}\right)  },\\
I_{2}\left(  \mathbf{R}\right)   &  =%
{\displaystyle\int}
\frac{d^{3}\mathbf{p}}{\left(  2\pi\right)  ^{3}}\frac{~e^{i\mathbf{p}%
\cdot\mathbf{R}}}{\left(  \mathbf{p}^{2}+M_{A}^{2}\right)  \left(
\mathbf{p}^{2}-m_{+}^{2}\right)  \left(  \mathbf{p}^{2}-m_{-}^{2}\right)  },\\
I_{3}\left(  \mathbf{R}\right)   &  =%
{\displaystyle\int}
\frac{d^{3}\mathbf{p}}{\left(  2\pi\right)  ^{3}}\frac{~e^{i\mathbf{p}%
\cdot\mathbf{R}}}{\left(  \mathbf{p}^{2}-m_{+}^{2}\right)  \left(
\mathbf{p}^{2}-m_{-}^{2}\right)  }.
\end{align}
where $\mathbf{R}=\left(  \mathbf{r}-\mathbf{r}^{\prime}\right)  $. In order
to solve these integrals, we first have factorized the denominator as
$\ \left(  \mathbf{p}^{2}+M_{A}^{2}\right)  ^{2}-\mathbf{v}_{0}^{2}%
\mathbf{p}^{2}=\left(  \mathbf{p}^{2}-m_{+}^{2}\right)  \left(  \mathbf{p}%
^{2}-m_{-}^{2}\right)  $. It is very important to remark that the massive
poles $m_{\pm}^{2},$
\begin{equation}
m_{\pm}^{2}=\frac{1}{2}\left[  \mathbf{v}_{0}^{2}-2M_{A}^{2}\pm\mathbf{v}%
_{0}\sqrt{\mathbf{v}_{0}^{2}-4M_{A}^{2}}\right]  ~,\ \ \ \label{poles1}%
\end{equation}
are positive under the condition $\mathbf{v}_{0}^{2}\geqslant4M_{A}^{2}$. This
fact is responsible for the oscillatory character of the magnetic sector
solutions to be achieved for this model. The remaining three integrals can be
solved in the complex plane, yielding:
\begin{align}
I_{1}\left(  \mathbf{R}\right)   &  =\frac{1}{4\pi R}~\left[  \frac{a_{+}}%
{a}\cos\left(  m_{+}R\right)  -\frac{a_{-}}{a}\cos\left(  m_{-}R\right)
\right]  ,\label{I1}\\
I_{2}\left(  \mathbf{R}\right)   &  =\frac{1}{4\pi R}~\left[  \frac
{e^{-M_{A}R}}{a_{+}a_{-}}~+\frac{\cos(m_{+}R)}{aa_{+}}-\frac{\cos(m_{-}%
R)}{aa_{-}}~\right]  ,\label{I2}\\
I_{3}\left(  \mathbf{R}\right)   &  =\frac{1}{4\pi R}~\frac{1}{a}\left[  \cos
m_{+}R-\cos m_{-}R\right]  , \label{I3}%
\end{align}
with
\begin{equation}
a=\mathbf{v}_{0}\sqrt{\mathbf{v}_{0}^{2}-4M_{A}^{2}},~~\ ~\ \ \ \ a_{\pm
}=\frac{1}{2}\left[  \mathbf{v}_{0}^{2}\pm\mathbf{v}_{0}\sqrt{\mathbf{v}%
_{0}^{2}-4M_{A}^{2}}\right]  .
\end{equation}
Replacing these results in eq. (\ref{G3}), the Green function $G_{jk}\left(
\mathbf{r}-\mathbf{r}^{\prime}\right)  $ is finally obtained:%
\begin{align}
G_{jk}\left(  \mathbf{R}\right)   &  =\frac{1}{4\pi R}\delta_{jk}\left[
\frac{a_{+}}{a}\cos\left(  m_{+}R\right)  -\frac{a_{-}}{a}\cos\left(
m_{-}R\right)  \right] \nonumber\\
&  -\frac{1}{4\pi R^{3}}~\frac{e^{-M_{A}R}}{M_{A}^{2}}\left[  \Delta
_{jk}(1+RM_{A})-M_{A}^{2}R_{j}R_{k}\right] \nonumber\\
&  ~-~\frac{1}{4\pi R^{3}}\mathbf{v}_{0}^{2}\left[  \frac{\left(  \Delta
_{jk}+m_{+}^{2}R_{j}R_{k}\right)  }{aa_{+}}\cos m_{+}R-\frac{\left(
\Delta_{jk}+m_{-}^{2}R_{j}R_{k}\right)  }{aa_{-}}~\cos m_{-}R\right]
\nonumber\\
&  -\frac{1}{4\pi R^{2}}\mathbf{v}_{0}^{2}~\Delta_{jk}\left[  \frac{m_{+}%
}{aa_{+}}\sin m_{+}R-\frac{m_{-}}{aa_{-}}\sin m_{-}R~\right] \nonumber\\
&  +~\frac{\mathbf{v}_{0}\epsilon_{jkl}R_{l}}{aR^{3}}\left[  \left(  \cos
m_{+}R-\cos m_{-}R\right)  +R\left(  m_{+}\sin m_{+}R-m_{-}\sin m_{-}R\right)
\right]  .
\end{align}
where $\Delta_{jk}=\left(  \delta_{kj}-3R_{j}R_{k}/R^{2}\right)  $. This Green
function can be simplified to its MCFJ counterpart taking the limit $\left(
M_{A}\rightarrow0\right)  $:%
\begin{align}
G_{jk}\left(  \mathbf{R}\right)   &  =\frac{1}{4\pi R}\delta_{jk}\cos\left(
\mathbf{v}_{0}R\right)  +\frac{1}{4\pi R^{3}}\Delta_{jk}\left[  \frac{R^{2}%
}{2}-\frac{R}{\mathbf{v}_{0}}\sin\left(  \mathbf{v}_{0}R\right)  +\frac
{1-\cos\left(  \mathbf{v}_{0}R\right)  }{\mathbf{v}_{0}^{2}}\right] \\
&  +\frac{1}{4\pi R^{3}}R_{j}R_{k}\left[  1-\cos\left(  \mathbf{v}%
_{0}R\right)  \right]  +~\frac{\epsilon_{jkl}R_{l}}{R^{3}}\left[
R\sin\mathbf{v}_{0}R-\frac{1-\cos\mathbf{v}_{0}R}{\mathbf{v}_{0}}\right]
.\nonumber
\end{align}
It is easy to verify that such result in the Lorentz symmetric limit
($\mathbf{v}_{0}\rightarrow0)$ rescues the pure Maxwell result%

\begin{equation}
G_{jk}\left(  \mathbf{R}\right)  =\frac{1}{4\pi R}\delta_{jk}.
\end{equation}

Turning back to the issue of calculating explicit classical solutions, we
address the solution for the scalar potential. Regarding eq. (\ref{A_pot}),
and the density current for a point-like charge, $J_{0}\left(  \mathbf{r}%
\right)  =e\delta\left(  \mathbf{r}\right)  $, the following expression is
obtained for the scalar potential
\begin{equation}
A_{0}\left(  \mathbf{r}\right)  =-\frac{e}{4\pi}\frac{e^{-M_{A}r}}{r}.
\label{A_zero1}%
\end{equation}
where $r=\left\vert \left\vert \mathbf{r}\right\vert \right\vert $. The
electric field may be easily evaluated from the scalar potential
($\mathbf{E}=-\nabla A_{0}),$ exhibiting an exponentially decaying solution as well:%

\begin{equation}
\mathbf{E}(\mathbf{r})=-\frac{e}{4\pi}\left(  \frac{M_{A}}{~r}+\frac{1}{r^{2}%
}\right)  e^{-M_{A}r}~\mathbf{\hat{r}}. \label{Electric1}%
\end{equation}
It is interesting to note that this is the same result obtained for the
electric field of the Maxwell-Proca Lagrangian (without Lorentz violation,
$V^{\mu}=0$). This means that the Lorentz-violating background, when coupled
to the gauge field as in Lagrangian (\ref{L1}), does not alter the
electrostatic sector. Analysis of Maxwell equations (\ref{M1},\ref{M2}%
,\ref{M3}) in the static regime reveals that this scenario remains true even
in the presence of a non-null current. Hence, the scalar potential and
electric field achieved here are the same for a point-like charge in uniform
motion (stationary solution), once the current $\mathbf{J}$ does not
contribute to $A_{0}$. In the absence of the Proca mass, the solutions
(\ref{A_zero1}, \ref{Electric1}) reduce to the CFJ ones, which coincide with
the Coulombian result:
\begin{equation}
A_{0}\left(  \mathbf{r}\right)  =-\frac{e}{4\pi}\frac{1}{r},\text{
\ \ \ }\mathbf{E}(\mathbf{r})=-\frac{e}{4\pi}\frac{1}{r^{2}}~\mathbf{\hat{r}}.
\end{equation}
The fixed background, therefore, does to not induce any effect on the electric
field solution of the MCFJ model, as well. This fact deserves to be compared
with the scenario of the dimensionally reduced version of this model, studied
in ref. \cite{Manojr1}. In such work, it was shown that the purely timelike
background alters the character of the classical Maxwell-Chern-Simons electric
solution, turning the usual screened Bessel-like solution into an unscreened
$1/r$ electric field.

The solution for vector potential can be found by the same procedure. From the
inverse tensor $\left[  D_{\nu\beta}\right]  ^{-1}$, we obtain the expression
for the vector potential Fourier transform:%

\begin{equation}
\mathbf{\tilde{A}}_{i}\left(  \mathbf{p}\right)  =-\frac{\left(
\mathbf{p}^{2}+M_{A}^{2}\right)  \tilde{J}_{i}+i\mathbf{v}_{0}\left(
\mathbf{p}\times\mathbf{\tilde{J}}\right)  _{i}}{[\left(  \mathbf{p}^{2}%
+M_{A}^{2}\right)  ^{2}-\mathbf{v}_{0}^{2}\mathbf{p}^{2}]},
\label{A_transform1}%
\end{equation}
where it was considered that $\mathbf{\tilde{J}}\cdot\mathbf{p}=0$, as a
consequence of current conservation $(p_{\alpha}\tilde{J}^{\alpha}=0)$. In
this case, we take the current associated with a point-like charge in uniform
motion with velocity\textbf{\ }$\mathbf{u}$,
\begin{equation}
~\mathbf{J}\left(  \mathbf{r}\right)  =e\mathbf{u}\delta\left(  \mathbf{r}%
\right)  , \label{J1}%
\end{equation}
In the momentum space, $\mathbf{\tilde{J}}\left(  \mathbf{p}\right)
\mathbf{=}e\mathbf{u}$. The vector potential, written as the Fourier transform
of eq. (\ref{A_transform1}), may be compactly expressed in terms of the
integrals (\ref{I1},\ref{I3}), so that:%
\begin{equation}
\mathbf{A}\left(  \mathbf{r}\right)  =-e\mathbf{u}I_{1}\left(  r\right)
+e\mathbf{v}_{0}~\mathbf{u}\times\mathbf{\nabla}I_{3}\left(  r\right)  .
\end{equation}
Considering the results already obtained, the vector potential takes the
form:
\begin{align}
\mathbf{A}\left(  \mathbf{r}\right)   &  =-\frac{e}{4\pi}\biggl\{\frac{1}%
{r}\left[  A_{+}\cos\left(  m_{+}r\right)  -A_{-}\cos\left(  m_{-}r\right)
\right]  \ \mathbf{u}\nonumber\\
&  +~\frac{\mathbf{v}_{0}}{ar^{3}}\left[  \left(  \cos m_{+}r-\cos
m_{-}r\right)  +r\left(  m_{+}\sin m_{+}r-m_{-}\sin m_{-}r\right)  \right]
~\mathbf{u}\times\mathbf{r}~\biggr\}, \label{A_MCFJP}%
\end{align}
where $A_{\pm}=a_{\pm}/a$. For the case of a vanishing Proca mass
($M_{A}\rightarrow0)$, this expression is reduced to the MCFJ vector
solution:
\begin{equation}
\mathbf{A}\left(  \mathbf{r}\right)  =-\frac{e}{4\pi}\biggl\{\frac{\cos\left(
\mathbf{v}_{0}r\right)  }{r}\mathbf{u}+\frac{1}{r^{3}}~\left[  \frac
{\cos(\mathbf{v}_{0}r)-1}{\mathbf{v}_{0}}+r\sin(\mathbf{v}_{0}r)\right]
\mathbf{u}\times\mathbf{r}\biggr\}. \label{A_CFJ}%
\end{equation}
It is interesting to note that oscillating solutions are obtained in both
cases, and the Proca mass is not a factor able to annihilate such a behavior.
For the case of a point-like static charge, the potential vector and the
magnetic field strength are null, once it depends only on the current source
$\left(  \mathbf{j}\neq0\right)  $. Hence, for a point-like charge in uniform
motion, $~\mathbf{J}\left(  \mathbf{r}\right)  =e\mathbf{u}\delta\left(
\mathbf{r}\right)  ,$ the magnetic field becomes non-null. It can be derived
from the vector potential $\left(  \mathbf{B}=\nabla\times\mathbf{A}\right)
$, yielding:
\begin{align}
\mathbf{B}\left(  \mathbf{r}\right)   &  =-\frac{e}{4\pi r^{3}}\biggl\{\left[
\!A_{+}\cos\left(  m_{+}r\right)  -A_{-}\cos\left(  m_{-}r\right)  +A_{+}%
m_{+}r\sin\left(  m_{+}r\right)  -A_{-}m_{-}r\sin\left(  m_{-}r\right)
\right]  (\mathbf{u}\times\mathbf{r)}\nonumber\\[0.08in]
&  -\frac{\mathbf{v}_{0}}{a}\left[  \!(1-m_{+}^{2}r^{2})\cos m_{+}%
r-(1-m_{-}^{2}r^{2})\cos m_{-}r+m_{+}r\sin\left(  m_{+}r\right)  -m_{-}%
r\sin\left(  m_{-}r\right)  \right]  \mathbf{u}\biggr\}. \label{B_MCFJP}%
\end{align}
An additional contribution proportional to $\left(  \mathbf{u\cdot r}\right)
\mathbf{r}$ also appears in the above expression, but it is was discarded
because the scalar product $\mathbf{u\cdot r}$ is null. This result stems from
the condition $\nabla\cdot\mathbf{A=0,}$ a consequence of the external current
conservation in the stationary regime. This magnetic field solution exhibits
two components: one in the direction the velocity $\mathbf{u}$, and another
orthogonal to the plane defined by $\mathbf{u}$ and $\mathbf{r.}$ This
magnetic field exhibits a decaying $1/r^{2}$ behavior both near the origin
($r\rightarrow0)$ and asymptotically. In the limit of a vanishing Proca mass
($M_{A}\rightarrow0),$ this expression is reduced to the MCFJ solution:%
\begin{equation}
\mathbf{B}\left(  \mathbf{r}\right)  =-\frac{e}{4\pi r^{3}}\biggl\{\left[
\cos(\mathbf{v}_{0}r)+\mathbf{v}_{0}r\sin(\mathbf{v}_{0}r)\right]
(\mathbf{u}\times\mathbf{r)}-\left[  \frac{1-\mathbf{v}_{0}^{2}r^{2}%
}{\mathbf{v}_{0}}\cos(\mathbf{v}_{0}r)-\frac{1}{\mathbf{v}_{0}}+r\sin
(\mathbf{v}_{0}r)\right]  \mathbf{u}\biggr\}. \label{B_CFJ}%
\end{equation}
This solution, similarly to the MCFJ-Proca one, is also decomposed in terms of
two orthogonal directions, $\mathbf{u}$ and $\mathbf{u}\times\mathbf{r}%
$\textbf{.} The fact that the MCFJ\ magnetic field exhibits an oscillating
behavior, associated with the $1/r^{2}$ behavior of the MCFJ electric field,
is compatible with the emission of Cerenkov radiation by a point-like charge
in uniform motion \cite{Cerenkov}. At the same way, the exponentially decaying
behavior of eq. (\ref{Electric1}) puts in evidence that no Cerenkov radiation
can be emitted by a stationary charge in the framework of the MCFJ-Proca
electrodynamics, once one condition to have radiation is that the fields
should present a non null asymptotic behavior \cite{Cerenkov}.

Another regime in which such solutions shall be investigated is the one of a
vanishing LV background ($\mathbf{v}_{0}\rightarrow0),$ which should lead back
to the usual Maxwell-Proca electrodynamics. {}This limit, however, can not be
implemented directly on the MCFJ-Proca expressions of eqs. (\ref{A_MCFJP}%
,\ref{B_MCFJP}), since these solutions were derived under the condition
$\mathbf{v}_{0}^{2}\geqslant4M_{A}^{2}$, which assures that the poles
(\ref{poles1}) be real and positive definite. A way to avoid this complication
is to implement the limit $\mathbf{v}_{0}\rightarrow0$ on eqs. (\ref{A_CFJ}%
,\ref{B_CFJ}). The results obtained,
\begin{equation}
\mathbf{A}\left(  \mathbf{r}\right)  =-\frac{e}{4\pi r}~\mathbf{u,}%
\ ~\ \ \mathbf{B}\left(  \mathbf{r}\right)  =-\frac{e}{4\pi r^{3}}%
~\mathbf{u}\times\mathbf{r}, \label{Proca}%
\end{equation}
nonetheless, do not recover the correct Maxwell-Proca behavior,
\begin{equation}
\mathbf{A}\left(  \mathbf{r}\right)  =-\frac{e}{4\pi r}e^{-M_{A}r}%
~\mathbf{u,}\ ~\ \ \mathbf{B}\left(  \mathbf{r}\right)  =-\frac{e}{4\pi r^{3}%
}\left[  1+M_{A}r\right]  e^{-M_{A}r}~\mathbf{u}\times\mathbf{r},
\label{ProcaEB}%
\end{equation}
attained from $\tilde{\mathbf{A}}_{i}\left(  p\right)  =-\tilde{J}_{i}/\left(
\mathbf{p}^{2}+M_{A}^{2}\right)  $ [eq. (\ref{A_transform1}) taken in the
limit\ $\mathbf{v}_{0}\rightarrow0$]. This apparent non equivalence simply
shows that Maxwell-Proca behavior may not be found as the limit $\mathbf{v}%
_{0}\rightarrow0\mathbf{\ }$of the MCFJ-Proca solutions. This is ascribed to
the structure of poles appearing in the MCFJ\ model, $1/\left(  \mathbf{p}%
^{2}+m^{2}\right)  $, associated with an exponentially decaying behavior, in
contrast with the MCFJ-Proca pole structure, $1/\left(  \mathbf{p}^{2}%
-m^{2}\right)  ,$ related to an oscillating behavior. In this way, we see that
the background turns the exponentially decaying behavior of the Maxwell-Proca
model into a oscillating solution that goes as $1/r^{2}$ far from the origin.
This is true in both MCFJ and MCFJ-Proca models.

\medskip

\begin{figure}[h]
\begin{center}
\scalebox{0.4}{\includegraphics{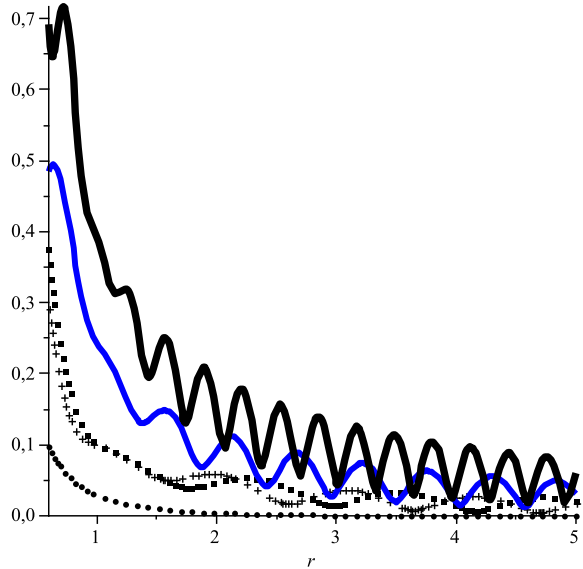}}
\end{center}
\caption{Simultaneous plot for the magnetic field in the radial interval
$0.4<r<4$ of the following models: Maxwell-Proca (for $M_{A}=1)$ - circle
dotted line; MCFJ - cross dotted line (for v$_{0}=3,M_{A}=0)$; MCFJ-Proca -
box dotted line (for v$_{0}=3,M_{A}=1)$, MCFJ-Proca - intermediary continuous
line (for v$_{0}=6,M_{A}=1)$ , MCFJ-Proca - thicker continuous line (for
v$_{0}=10,M_{A}=1),$ and $u=0.5$ in all cases.}%
\end{figure}

The graph of Fig.1 depicts a simultaneous plot of the Maxwell-Proca, MCFJ and
MCFJ-Proca magnetic fields in the case the velocity $\mathbf{u}$ is orthogonal
to the vector $\mathbf{r}$. Such a graph shows clearly that the MCFJ-Proca
solution deviates substantially from the Maxwell-Proca behavior mainly due to
the presence of oscillation. This deviation increases with the background
magnitude: the larger is the background, more pronounced is the amplitude and
the frequency of such oscillations. For a background not so large in
comparison with the Proca mass $\left(  \mathbf{v}_{0}\gtrsim M_{A}\right)  $,
the MCFJ e MCFJ-Proca solutions are close from each other. Keeping $M_{A}$
constant while $\mathbf{v}_{0}$ increases, it occurs that these solutions
become different. This behavior is confirmed in the graph of Fig.2, where it
is shown a comparison between the MCFJ and MCFJ-Proca magnetic solutions for
equal values of $\mathbf{v}_{0}$ (for the case the velocity $\mathbf{u}$ is
orthogonal to the vector $\mathbf{r).}$ In general, these solutions are nearly
coincident, but deviate from each other for larger $M_{A}$
magnitudes.\begin{figure}[h]
\begin{center}
\scalebox{0.4}{\includegraphics{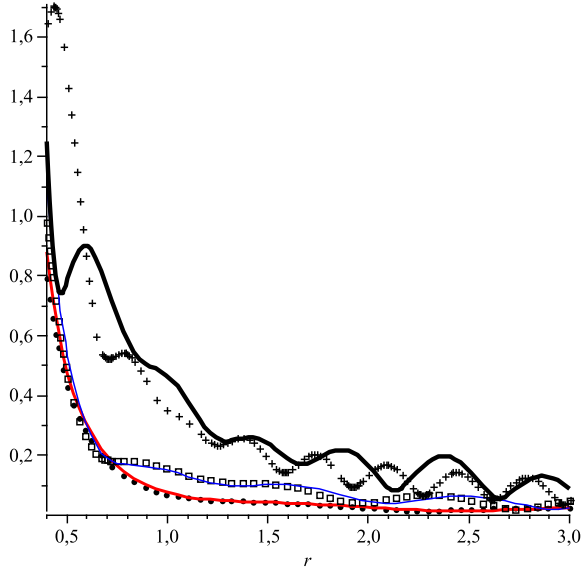}}
\end{center}
\caption{Simultaneous plot of the magnetic fields for the MCFJ (dotted curves
- $M_{A}=0$) and MCFJ-Proca (continuous curves) models in the radial interval
$0.4<r<3$. The lower two curves - point dotted and continuous superposed line
- correspond to (v$_{0}=2,M_{A}=0)$ and (v$_{0}=2,M_{A}=0.5).$ For the two
intermediary curves - box dotted and continuous thin line: (v$_{0}%
=4,M_{A}=0,)$ and (v$_{0}=4,M_{A}=1).$ For the two upper curves - cross dotted
and thicker continuous line: (v$_{0}=9,M_{A}=0)$ and (v$_{0}=9,M_{A}=4).$ It
was used $u=0.5$ in all graphs.}%
\end{figure}

In order to find a general expression for the magnetic field strength, a
direct way consists in starting from wave equation (\ref{B1}), which in the
stationary regime and for a purely timelike background is simplified to:
\begin{equation}
\left[  \mathbf{\nabla}^{2}(\mathbf{\nabla}^{2}-2M_{A}^{2}+\mathbf{v}_{0}%
^{2})+M_{A}^{4}\right]  \mathbf{B=-}M_{A}^{2}\left(  \mathbf{\nabla}%
\times\mathbf{j}\right)  -\mathbf{v}_{0}\mathbf{\nabla}^{2}%
\mathbf{j+\mathbf{\mathbf{\nabla}^{2}}\left(  \mathbf{\nabla}\times
\mathbf{j}\right)  .}%
\end{equation}
The corresponding Green function for this equation is $G\left(  \mathbf{r}%
-\mathbf{r}^{\prime}\right)  =\!\!\displaystyle\int\frac{d^{3}\mathbf{p}%
}{\left(  2\pi\right)  ^{3}}\,\tilde{G}\left(  \mathbf{p}\right)
e^{i\mathbf{p}\cdot\left(  \mathbf{r}-\mathbf{r}^{\prime}\right)  },$ with
$\tilde{G}\left(  p\right)  =[\left(  \mathbf{p}^{2}-m_{+}^{2}\right)  \left(
\mathbf{p}^{2}-m_{-}^{2}\right)  ]^{-1},$ and the poles $m_{\pm}^{2}$ being
given by eq. (\ref{poles1}). In true, this Green function was already
evaluated, corresponding to the result of integral $I_{3}$, so that the
magnetic field strength is:
\begin{equation}
\mathbf{B}\left(  \mathbf{r}\right)  =\frac{1}{4\pi a}~\int\frac{1}{R}\left[
\cos m_{+}R-\cos m_{-}R\right]  [-M_{A}^{2}\mathbf{\nabla}\times
\mathbf{j}\left(  \mathbf{r}^{\prime}\right)  -\mathbf{v}_{0}\mathbf{\nabla
}^{2}\mathbf{j\left(  \mathbf{r}^{\prime}\right)  +\mathbf{\nabla}^{2}\nabla
}\times\mathbf{j}\left(  \mathbf{r}^{\prime}\right)  ]d^{3}\mathbf{r}^{\prime
}.
\end{equation}
This outcome differs from the MCFJ one presented in ref. \cite{CFJ}, which
exhibits an exponentially decaying behavior. For the case of the pure CFJ
model ($M_{A}\rightarrow0)$, we get:
\begin{equation}
\mathbf{B}\left(  \mathbf{r}\right)  =\frac{1}{4\pi\mathbf{v}_{0}^{2}}~\int
d^{3}\mathbf{r}\left(  \frac{1-\cos(\mathbf{v}_{0}R)}{R\mathbf{\ }}\right)
\mathbf{\nabla}^{2}\left[  \mathbf{v}_{0}\mathbf{j\left(  \mathbf{r}^{\prime
}\right)  -\nabla}\times\mathbf{j}\left(  \mathbf{r}^{\prime}\right)  \right]
~,\label{B6}%
\end{equation}
with $R=\left\vert \left\vert \mathbf{r-r}^{\prime}\right\vert \right\vert $.
Using the Green theorem, $\int_{V}\left[  \phi\mathbf{\nabla}^{2}\chi
-\chi\mathbf{\nabla}^{2}\phi\right]  d^{3}r=\int_{\partial R}\left[
\phi\partial\chi/\partial n-\chi\partial\phi/\partial n\right]  dS$, and
considering that the current and its derivatives are null on a very distant
surface, then eq. (\ref{B6}) is rewritten as:
\begin{equation}
\mathbf{B}\left(  \mathbf{r}\right)  =\frac{1}{4\pi}~\int\left(  \frac
{\cos(\mathbf{v}_{0}R)}{R\mathbf{\ }}\right)  \left[  \mathbf{v}%
_{0}\mathbf{j\left(  \mathbf{r}^{\prime}\right)  -\nabla}\times\mathbf{j}%
\left(  \mathbf{r}^{\prime}\right)  \right]  ~d^{3}\mathbf{r.}%
\end{equation}
where it was used $\nabla^{2}[(1-\cos($v$_{0}R))/R]=\mathbf{v}_{0}^{2}%
\cos(\mathbf{v}_{0}R)/R$. This result is equal to the one of ref. \cite{CFJ}
apart from a\textbf{ }global signal stemming from our definition of the
external current vector\textbf{ $\mathbf{j}\left(  \mathbf{r}\right)  $}.

\subsection{Solution for a purely spacelike background}

The case the background is purely spacelike, $V^{\alpha}=\left(
0,\mathbf{v}\right)  ,$ becomes particularly interesting when we consider the
model of Lagrangian (\ref{L1}) under the perspective of its physical
consistency. In fact, it is known that this model exhibits full consistency
(stability, causality, unitarity) only for a spacelike background. Henceforth,
this is a good reason to study the corresponding classical solutions for a
purely spacelike background. We begin writing the elements of the inverse
matrix (\ref{Mat1}) for $V^{\alpha}=\left(  0,\mathbf{v}\right)  $:
\begin{align}
\left[  D^{-1}\right]  _{00}  &  =\frac{p^{2}-M_{A}^{2}}{D}-\frac
{\mathbf{v}^{2}}{D\left(  p^{2}-M_{A}^{2}\right)  }\left(  p_{0}\right)
^{2},\\
\left[  D^{-1}\right]  _{0k}  &  =i\epsilon_{kjl}\mathbf{v}_{j}p_{l}%
+\frac{\left(  \mathbf{v\cdot p}\right)  \mathbf{v}_{k}-\mathbf{v}^{2}p_{k}%
}{D\left(  p^{2}-M_{A}^{2}\right)  }p_{0},\\
\left[  D^{-1}\right]  _{k0}  &  =-i\epsilon_{kjl}\mathbf{v}_{j}p_{l}%
+\frac{\left(  \mathbf{v\cdot p}\right)  \mathbf{v}_{k}-\mathbf{v}^{2}p_{k}%
}{D\left(  p^{2}-M_{A}^{2}\right)  }p_{0},\\
\left[  D^{-1}\right]  _{jk}  &  =-\frac{p^{2}-M_{A}^{2}}{D}\delta_{jk}%
-\frac{\mathbf{v}^{2}p_{j}p_{k}-{p}^{2}\mathbf{v}_{j}\mathbf{v}_{k}-\left(
\mathbf{v\cdot p}\right)  \left(  \mathbf{v}_{j}p_{k}+\mathbf{v}_{k}%
p_{j}\right)  }{D\left(  p^{2}-M_{A}^{2}\right)  }+\frac{i\epsilon
_{jkl}\mathbf{v}_{l}p_{0}}{D},
\end{align}
with $D=\left(  {p}^{2}-M_{A}^{2}\right)  ^{2}-\mathbf{v}^{2}p^{2}-\left(
\mathbf{v\cdot p}\right)  ^{2}$.

As we are to study the stationary solutions of the model, we set $p_{0}=0$ in
the above equations to get the Green function for the wave equation for the
stationary four potential:
\begin{align}
G_{00}\left(  \mathbf{p}\right)   &  =-\frac{\mathbf{p}^{2}+M_{A}^{2}}{R_{s}%
},~\ \ \ \ \ \ \ \ G_{0k}\left(  \mathbf{p}\right)  =i\frac{\left(
\mathbf{v}\times\mathbf{p}\right)  _{k}}{R_{s}}=-G_{k0},\label{Mat3}\\
G_{jk}\left(  \mathbf{p}\right)   &  =\frac{\mathbf{p}^{2}+M_{A}^{2}}{R_{s}%
}\delta_{jk}+\frac{\delta_{jk}\left(  \mathbf{v\times p}\right)  \cdot\left(
\mathbf{v\times p}\right)  -\left(  \mathbf{v\times p}\right)  _{j}\left(
\mathbf{v\times p}\right)  _{k}}{\left(  \mathbf{p}^{2}+M_{A}^{2}\right)
R_{s}},
\end{align}
where
\begin{equation}
R_{s}=\left(  \mathbf{p}^{2}+M_{A}^{2}\right)  ^{2}+\mathbf{p}^{2}%
\mathbf{v}^{2}-\left(  \mathbf{p}\cdot\mathbf{v}\right)  ^{2}.
\end{equation}
Supposing that $\theta$ is the angle between the vectors $\mathbf{p}$ and
$\mathbf{v}$, ($\mathbf{p}\cdot\mathbf{v=|p||v|\cos}\theta)$, the denominator
$R_{s}$ is read as:%
\begin{equation}
R_{s}=\left(  \mathbf{p}^{2}+M_{A}^{2}\right)  ^{2}+\mathbf{p}^{2}%
\mathbf{v}^{2}\sin^{2}\theta.
\end{equation}
Similarly to the timelike case, we can construct the stationary Green function
from the inverse matrix (\ref{Mat3}). In this sense, we write:%
\begin{align}
G_{00}\left(  \mathbf{r}-\mathbf{r}^{\prime}\right)   &  =-F_{2}\left(
\mathbf{r}-\mathbf{r}^{\prime}\right)  ,\text{ \ \ }G_{0k}\left(
\mathbf{r}-\mathbf{r}^{\prime}\right)  =(\mathbf{v}\times\mathbf{\nabla)}%
_{k}~F_{1}\left(  \mathbf{r}-\mathbf{r}^{\prime}\right)  ,\\
G_{jk}\left(  \mathbf{r}-\mathbf{r}^{\prime}\right)   &  =\delta_{jk}%
F_{2}\left(  \mathbf{r}-\mathbf{r}^{\prime}\right)  -\left[  \delta
_{jk}\left(  \mathbf{v\times\nabla}\right)  \cdot\left(  \mathbf{v\times
\nabla}\right)  -\left(  \mathbf{v\times\nabla}\right)  _{j}\left(
\mathbf{v\times\nabla}\right)  _{k}\right]  F_{3}\left(  \mathbf{r}%
-\mathbf{r}^{\prime}\right)
\end{align}
where the Fourier transforms $F_{i}$ are defined as:
\begin{align}
F_{1}\left(  \mathbf{R}\right)   &  =%
{\displaystyle\int}
\frac{d^{3}\mathbf{p}}{\left(  2\pi\right)  ^{3}}\frac{~e^{i\mathbf{p}%
\cdot\mathbf{R}}}{R_{s}},\label{F1}\\
F_{2}\left(  \mathbf{R}\right)   &  =%
{\displaystyle\int}
\frac{d^{3}\mathbf{p}}{\left(  2\pi\right)  ^{3}}\frac{\left(  \mathbf{p}%
^{2}+M_{A}^{2}\right)  ~e^{i\mathbf{p}\cdot\mathbf{R}}}{R_{s}},\label{F2}\\
F_{3}\left(  \mathbf{R}\right)   &  =%
{\displaystyle\int}
\frac{d^{3}\mathbf{p}}{\left(  2\pi\right)  ^{3}}\frac{~e^{i\mathbf{p}%
\cdot\mathbf{R}}}{\left(  \mathbf{p}^{2}+M_{A}^{2}\right)  R_{s}}. \label{F3}%
\end{align}
The main difficult concerning the Fourier transforms (\ref{F1}-\ref{F3}) is
that we can not calculate an exact solution for such integrations due to the
presence of the angular factor in the denominator $R_{s}$. A possible way to
circumvent this impossibility is to perform the integration under some
approximation. A feasible approximation consists in considering the background
small before the Proca mass ($\mathbf{v}^{2}\ll M_{A}^{2}),$ which implies the
following second order expansion:
\begin{equation}
\frac{1}{R_{s}}=\frac{1}{\left(  \mathbf{p}^{2}+M_{A}^{2}\right)  ^{2}}%
-\frac{\mathbf{p}^{2}\sin^{2}\theta}{\left(  \mathbf{p}^{2}+M_{A}^{2}\right)
^{4}}\mathbf{v}^{2}+\mathcal{O}\left(  \text{v}^{4}\right)  .
\end{equation}
With it, and regarding the special situation in which the background
$\mathbf{v}$ and the vector $\mathbf{r}$ are parallel, the following second
order solutions are obtained:
\begin{align}
F_{1}\left(  R\right)   &  =\frac{1}{8\pi M_{A}}\left[  1-\frac{\mathbf{v}%
^{2}}{12}\left(  \frac{R}{M_{A}}+\frac{1}{M_{A}^{2}}\right)  \right]
e^{-M_{A}R}~\ ,\label{F1A}\\[0.2cm]
F_{2}\left(  R\right)   &  =\frac{1}{4\pi}\left[  \frac{1}{R}-\frac
{\mathbf{v}^{2}}{4M_{A}}\right]  e^{-M_{A}R},\label{F1AB}\\[0.2cm]
F_{3}\left(  R\right)   &  =\frac{1}{32\pi}\left[  ~\frac{1}{M_{A}^{3}}%
+\frac{R}{M_{A}^{2}}-\frac{\mathbf{v}^{2}}{24}\left(  \frac{3}{M_{A}^{5}%
}+\frac{3R}{M_{A}^{4}}+\frac{R^{2}}{M_{A}^{3}}\right)  \right]  e^{-M_{A}R}
\label{F1B}%
\end{align}
The Green functions are finally written as:%
\begin{align}
G_{00}\left(  \mathbf{R}\right)   &  =-\frac{1}{4\pi}\left[  \frac{1}{R}%
-\frac{\mathbf{v}^{2}}{4M_{A}}\right]  e^{-M_{A}R},\\
G_{0i}\left(  \mathbf{R}\right)   &  =\frac{1}{8\pi}\left[  -\frac{1}{R}%
+\frac{\mathbf{v}^{2}}{12M_{A}}\right]  e^{-M_{A}R}\left(  \mathbf{v}%
\times\mathbf{R}\right)  _{i}\\
G_{jk}\left(  \mathbf{R}\right)   &  =\frac{1}{4\pi}\biggl\{\delta_{jk}\left[
\frac{1}{R}-\frac{\mathbf{v}^{2}}{8}\left(  R+\frac{1}{M_{A}}\right)
+\frac{1}{8R}\left(  \mathbf{v}\cdot\mathbf{R}\right)  ^{2}\right]
e^{-M_{A}R}\\
&  +\frac{1}{8}\left[  \frac{\mathbf{v}_{j}\mathbf{v}_{k}}{M_{A}}+\frac{1}%
{R}\left(  \mathbf{v\times R}\right)  _{j}\left(  \mathbf{v\times R}\right)
_{k}\right]  e^{-M_{A}R}+\mathcal{O}\left(  \mathbf{v}^{4}\right)
\biggr\}\nonumber
\end{align}

We can now write the solutions for the scalar and vector potential, starting
from the corresponding Fourier transforms extracted from eqs. (\ref{A1}%
,\ref{Mat3}):
\begin{align}
\tilde{A}_{0}\mathbf{(p)}  &  =-\frac{\left(  \mathbf{p}^{2}+M_{A}^{2}\right)
}{R_{s}}\tilde{J}_{0}+\frac{i\left(  \mathbf{p}\times\mathbf{v}\right)
\cdot\mathbf{\tilde{J}}}{R_{s}}\label{A01}\\[0.2cm]
\mathbf{\tilde{A}(p)}  &  =\frac{i\left(  \mathbf{p}\times\mathbf{v}\right)
}{R_{s}}\tilde{J}_{0}\text{\ }-\frac{\left(  \mathbf{p}^{2}+M_{A}^{2}\right)
\mathbf{\tilde{J}}}{R_{s}}-\frac{\left(  \mathbf{p}\times\mathbf{v}\right)
\cdot\left(  \mathbf{p}\times\mathbf{v}\right)  \mathbf{\tilde{J}}}{\left(
\mathbf{p}^{2}+M_{A}^{2}\right)  ~R_{s}}+\left(  \mathbf{p}\times
\mathbf{v}\right)  \frac{\left(  \mathbf{p}\times\mathbf{v}\right)
\cdot\mathbf{\tilde{J}}}{\left(  \mathbf{p}^{2}+M_{A}^{2}\right)  ~R_{s}}.
\label{AV1}%
\end{align}
Such equations show clearly that the electric field and the magnetic sectors
are now entwined. In fact, a static charge is able to create a magnetic field
$\left(  \mathbf{\tilde{A}\neq0}\right)  $ as much a stationary current is
capable to imply a non-null electric field $\left(  \mathbf{A}_{0}%
\mathbf{\neq0}\right)  $. Hence, the electric and magnetic fields coexist
simultaneously both for a static charge or a stationary current. Such scenario
occurs only for a space-like Lorentz breaking background. We now study the
solutions corresponding to a point-like charge [$J_{0}\left(  \mathbf{r}%
\right)  =e\delta\left(  \mathbf{r}\right)  ,\ \mathbf{J}\left(
\mathbf{r}\right)  =e\mathbf{u}\delta\left(  \mathbf{r}\right)  ]$. In
momentum space, $\tilde{J}_{0}\left(  p\right)  =e$, $\tilde{J}\left(
p\right)  =e\mathbf{u.}$ Working at second order in the background magnitude,
we obtain the following expressions for scalar and vector potential as Fourier
transform of the expressions (\ref{A01},\ref{AV1}), so that it holds:
\begin{align}
A_{0}\mathbf{(r)}  &  =-eF_{2}\left(  r\right)  +e\left[  \left(
\mathbf{\nabla}\times\mathbf{v}\right)  \cdot\mathbf{u}\right]  F_{1}\left(
r\right)  ,\\[0.2cm]
\mathbf{A(r)}  &  =e\left(  \mathbf{\nabla}\times\mathbf{v}\right)
F_{1}\left(  r\right)  \text{\ }-eF_{2}\left(  r\right)  \text{\ }%
\mathbf{u}+e\left[  (\mathbf{\nabla}\times\mathbf{v)\cdot}(\mathbf{\nabla
}\times\mathbf{v)u-}\left(  \mathbf{\nabla}\times\mathbf{v}\right)  [\left(
\mathbf{\nabla}\times\mathbf{v}\right)  \cdot\mathbf{u]}\right]  F_{3}\left(
r\right)  \text{\ }.
\end{align}

Taking the expressions (\ref{F1A},\ref{F1B}), the solutions for these
potentials are readily achieved:%
\begin{align}
A_{0}(\mathbf{r}) &  =-\frac{e}{4\pi}\left(  \frac{1}{r}-\frac{\mathbf{v}^{2}%
}{4M_{A}}\right)  e^{-M_{A}r}\ ,\\[0.2cm]
\mathbf{A}(\mathbf{r}) &  =-\frac{e}{4\pi}\left[  \left(  \frac{1}{r}%
-\frac{\mathbf{v}^{2}}{8M_{A}}\right)  \mathbf{u+}\frac{\mathbf{1}}{8M_{A}%
}\left(  \mathbf{v}\cdot\mathbf{u}\right)  \mathbf{v}\right]  e^{-M_{A}r},
\end{align}
These are the general expressions for the scalar and vector potential in
$\mathbf{v}^{2}$ order taking into account the condition that~$\mathbf{v}$ and
$\mathbf{r}$ are parallel. It is seen that the vector potential only retain
the contributions proportional to the velocity $\mathbf{u}$, while these
contributions do not appear in the scalar potential. This means that the
current does not contribute to the electric sector and the static charge does
not yield magnetic field. The explicit expressions for the electric and
magnetic field strength,%
\begin{align}
\mathbf{E}(\mathbf{r}) &  =-\frac{e}{4\pi}\left(  \frac{1}{r^{2}}+\frac{M_{A}%
}{r}-\frac{\mathbf{v}^{2}}{4}\right)  e^{-M_{A}r}~\mathbf{\hat{r}}.\\[0.2cm]
\mathbf{B}(\mathbf{r}) &  =-\frac{e}{4\pi}\left[  \frac{1}{r^{2}}+\frac{M_{A}%
}{r}-\frac{\mathbf{v}^{2}}{8}\right]  e^{-M_{A}r}~\mathbf{u\times\hat{r}}.
\end{align}
confirm that this is really the case (at least in the context of the
approximations adopted) in $\mathbf{v}^{2}$ order taking into account the
condition that~$\mathbf{v}$ and $\mathbf{r}$ are parallel. Near the origin,
these solutions present a $1/r^{2}$ behavior, while far from the origin both
fields possess a totally screened behavior, which avoids any attempt to obtain
Cerenkov radiation for such background. It is instructive to mention that
these expressions provide the Maxwell-Proca solutions in the limit $\left(
\mathbf{v\rightarrow0}\right)  $, given by eqs. (\ref{Electric1}%
,\ref{ProcaEB}). Although the limit $\left(  M_{A}\rightarrow0\right)  $ could
be easily carried out in the above expressions, such a limit has no sense
here, once these equations have been derived under the supposition that
($\mathbf{v}^{2}\ll M_{A}^{2}).$

\section{Conclusion}

In this work, we have studied the classical static and stationary solutions of
the Lorentz-violating MCFJ-Proca model, while the MCFJ solutions were also
obtained as the limit of a vanishing Proca mass $\left(  M_{A}%
\mathbf{\rightarrow0}\right)  $. Starting from the equation of motion for the
four-potential, the corresponding Green function was written in a matrix form,
which provides explicit expressions for the four-potential in momentum space.
In the purely timelike case, it was written the Green function for a
stationary configuration. An exponentially decaying solution was attained for
the electric sector, which is equal to the Maxwell-Proca solutions. In the
limit $\left(  M_{A}\mathbf{\rightarrow0}\right)  $, these solutions recover
the Coulombian ones, which reveals that the background does not modify the
electric sector of both MCFJ-Proca and MCFJ models. This is to be contrasted
with the LV low-dimensional scenario of the model of ref. \cite{Manojr1},
established as the dimensional reduction of the MCFJ model, where the
Maxwell-Chern-Simons electric sector was drastically altered by the presence
of the timelike background. In concerning the magnetic sector, the background
induces a radical modification. In fact, while the Lorentz symmetric
Maxwell-Proca model exhibits an exponentially decaying magnetic field
strength, the MCFJ-Proca model presents an oscillating magnetic field solution
for charges in stationary motion. This change in behavior is ascribed to the
sign reversion in the poles of the theory. Taking the limit $\left(
M_{A}\mathbf{\rightarrow0}\right)  ,$ the magnetic solution for a stationary
charge in the MCFJ model is obtained, whose oscillating behavior is compatible
with the emission of Cerenkov radiation in such framework, as shown in refs.
\cite{Cerenkov}. The emission of Cerenkov radiation in the context of the
MCFJ-Proca model is not allowed, due to the exponentially screened behavior of
the electric field.

In the purely spacelike case, the same procedure was employed. The Green
function for a stationary scenario was written in terms of Fourier integrals,
which were solved in a second order approximation into the regime of a small
background $\left(  \mathbf{v}^{2}\ll M_{A}^{2}\right)  $. The solutions were
derived for a special configuration in which the background is parallel to the
position of observation of the fields. It was then verified that the electric
and magnetic sectors present exponentially decaying solutions, which recover
the Maxwell-Proca behavior in the limit $\left(  \mathbf{v\rightarrow
0}\right)  $.

In a forthcoming work, we intend to use the Green function techniques to
investigate the classical radiation emitted by a point-like charge in uniform
linear acceleration and in uniform circular motion (synchrotron radiation). In
this case, we shall search for time-dependent solutions for the Green
functions which may lead to the calculation of the time-dependent solutions
that yield radiation.

\begin{acknowledgments}
The authors are grateful to FAPEMA (Funda\c{c}\~{a}o de Amparo \`{a} Pesquisa
do Estado do Maranh\~{a}o), to CNPq (Conselho Nacional de Desenvolvimento
Cient\'{\i}fico e Tecnol\'{o}gico), and CAPES for financial support.
\end{acknowledgments}

\end{document}